\documentclass[aps,prb,twocolumn,showpacs]{revtex4}
\usepackage{epsfig}
\usepackage{mathrsfs}
\usepackage{amssymb}
\usepackage{color}
\usepackage[toc,page]{appendix}

\newcommand{\bcen}{\begin{center}}
\newcommand{\ecen}{\end{center}}
\newcommand{\btab}{\begin{tabular}}
\newcommand{\etab}{\end{tabular}}
\newcommand{\bdes}{\begin{description}}
\newcommand{\edes}{\end{description}}

\newcommand{\beq}{\begin{equation}}
\newcommand{\eeq}{\end{equation}}
\newcommand{\bea}{\begin{eqnarray}}
\newcommand{\eea}{\end{eqnarray}}
\newcommand{\non}{\nonumber}

\newcommand{\bary}{\begin{array}}
\newcommand{\eary}{\end{array}}
\newcommand{\benum}{\begin{enumerate}}
\newcommand{\eenum}{\end{enumerate}}
\newcommand{\bitem}{\begin{itemize}}
\newcommand{\eitem}{\end{itemize}}

\newcommand{\ba} { \mbox{\boldmath $a$}}

\newcommand{\bk} { \mbox{\boldmath $k$}}

\newcommand{\bR} { \mbox{\boldmath $R$}}
\newcommand{\bS} { \mbox{\boldmath $S$}}

\newcommand{\mean}[1]{\langle #1 \rangle}
\newcommand{\bra}[1]{{\langle #1 |}}
\newcommand{\ket}[1]{| #1 \rangle}

\newcommand{\eqn}[1] {eqn.~(\ref{#1})}
\newcommand{\prn}[1] {(\ref{#1})}

\newcommand{\fig}[1]{fig.~\ref{#1}}
\newcommand{\Fig}[1]{Fig.~\ref{#1}}

\newcommand{\figwidth}{8.0truecm}

\newcommand{\citebyname}[1]{\citeauthor{#1}\cite{#1}}

\def\tc{$T_c~$}

\def\c2{ CuO$_2~$}

\def\ppi{$p\pi$~}

\def\he4{${\rm {}^4He}$~}

\begin{document}

\title{Possibility of High $T_c$ Superconductivity in doped Graphene}
\author{Sandeep Pathak$^{1,2}$, Vijay B.~Shenoy$^{2,1}$ and G. Baskaran$^3$}
\affiliation{$^1$Materials Research Centre, Indian Institute of Science, Bangalore 560012, India\\
$^2$Centre for Condensed Matter Theory, Department of Physics, Indian Institute of Science, Bangalore 560012, India\\
$^3$Institute of Mathematical Sciences, Chennai 600113, India}


\maketitle


{\bf Superconductivity at room temperatures, with its multifarious
technological possibilities, is a phenomenon that is yet to be realized
in a material system. The search for such systems has lead to the
discovery of ``high temperature'' superconducting materials such as
the fascinating cuprates,\cite{Bednorz1986}$^,$\cite{Anderson1987}  MgB$_2$\cite{Nagamatsu2001}
and most recently, Fe-based pnictides.\cite{Kamihara2008}
Graphene is a remarkable two dimensional
conductor.\cite{Geim2007,Katsnelson2007,CastroNeto2007} A newly discovered method to
cleave and isolate single or finite number of atomic layers of
graphene, its mechanical robustness and novel electrical properties
has caught the attention of the scientific and nanotechnology
communities. Undoped graphene is a semi-metal and does not superconduct at low temperatures. However, on ``doping optimally''
if  graphene supports high \tc superconductivity it will
make graphene even more valuable from basic science and technology
points of view. Here we build on a seven year old suggestion of
\citebyname{Baskaran2002} (GB) of an electron correlation based
mechanism of superconductivity for graphite like systems and
demonstrate theoretically the possibility of high temperature
superconductivity in {\it doped} graphene.}

GB\cite{Baskaran2002} suggested the possibility of high
temperature superconductivity in graphene based on an effective
phenomenological Hamiltonian that combined band theory and Pauling's
idea of resonating valance bonds (RVB).  The model predicted a
vanishing $T_c$ for undoped graphene, consistent with
experiments. However, for doped graphene superconducting estimates of
\tc's were embarrassingly high.  Very recently Black-Schaffer and Doniach\cite{Schaffer2007}
used GB's effective Hamiltonian and studied graphitic systems and
found that a superconducting state with $d + i d$ symmetry to be the
lowest energy state in a mean field theory. The mean field theory also
predicts a rather high value of the optimal $T_c$. Other authors have
studied possibility of superconductivity based on electron-electron
and electron-phonon interactions.\cite{Uchoa2007,Choy1995,Furukawa2001,Onari2003,CastroNeto2007}
 While there is an encouraging signal for high
$T_c$ superconductivity in the phenomenological GB model, it is
important to establish this possibility by the study of a more basic
and realistic model.  Since the motivation for GB model arose from a
repulsive Hubbard model, here we directly analyse this more basic
repulsive Hubbard model that describes low energy properties of graphene. We construct variational
wavefunctions  motivated by RVB physics, and perform extensive
Monte Carlo study incorporating  crucial correlation effects. \textit{This
approach which has proved to be especially successful in understanding
the ground state of cuprates, clearly points to a 
superconducting ground state in  doped graphene}. Further support
is obtained from a slave rotor analysis which also includes
correlation effects.  Our estimate of the Kosterlitz-Thouless
superconducting $T_c$ is of the order of room temperatures, and we also
discuss experimental observability of our prediction of high
temperature superconductivity in graphene.

Low energy electrical and magnetic properties of graphene are well described\cite{Hubbard1995}
by a tight binding Hubbard model defined on a honeycomb lattice with a single
2p$_z$ orbital per carbon atom:
\bea
{\cal H}_{H} = - \sum_{\langle ij\rangle}t^{}_{ij}  c^\dagger_{i\sigma} c^{}_{j\sigma} 
+ h.c. + U \sum_i n^{}_{i\uparrow}n^{}_{i\downarrow}
\eea 
Here, $i$ labels atomic sites, $c_{i \sigma}$ is an annihilation
operator for an electron with spin $\sigma$ at site $i$, $n_{i
  \sigma}$ is the number operator at site $i$ of $\sigma$ spin
electrons, $t$ $\approx$ 2.5 eV is the hopping matrix element and $U$
$\approx$ 6 eV is the onsite Hubbard repulsion. The unique band
structure of the above model leads to a `Dirac cone' type of spectrum
for electron motion close to two points in the Brillouin zone, giving
rise to a density of states that varies linearly with
energy near zero energy (half filling).

The Hubbard $U$ is about half the \ppi free bandwidth, and this places
graphene in an intermediate or weak coupling regime. Based on this one
is tempted to conclude that electron correlations are not
important. Nonetheless electron correlations are known to be important
in finite \ppi bonded planar molecular systems such as benzene,
naphthane, anthracene, caronene etc, all having nearly the same value
for quantum chemical parameters $t$ and $U$.\cite{Hubbard1995} One of
the consequences of this is that the first excited spin-0 state lies
above the first excited spin-1 state by more than 1 eV. There is a
predictable consequence of this large singlet-triplet splitting:
graphene can be viewed as an end member of a sequence of planar \ppi
bonded system; this has been also suggested to have a new spin-1
collective mode spectrum, as a consequence of finite
$U$.\cite{BaskaranJafari2002}

Pauling\cite{Pauling1960} was the first to recognize
dominance of singlet correlation between two neighboring \ppi
electrons in the ground state. He argued that doubly occupied or empty
2p$_z$ orbitals (polar configurations) are less important because of
electron-electron repulsion in the 2p$_z$ orbital. Pauling thus ignored polar
configurations. Once we ignore polar fluctuations (states with double
occupancy) and consider a resonance among the nearest neighbor valence
bond configurations we get the well known RVB state.  However, such a
Hilbert space actually describes a Mott insulating state rather than a
metal. Experimentally, undoped graphene is a broad band conductor,
albeit with a linearly vanishing density of states at the Fermi
energy.

To recover metallicity in Pauling's RVB theory, GB combined the
broad band feature of \ppi electrons with Pauling's real space singlet
(covalent) bonding tendency and suggested\cite{Baskaran2002} a low energy phenomenological
model for graphene:
\bea
{\cal H}_{\rm GB} = - \sum_{\langle ij\rangle}t^{}_{ij}  c^\dagger_{i\sigma} c^{}_{j\sigma} 
+ h.c.  - J\sum_{\langle ij\rangle} {\bf b}^{\dagger}_{ij} {\bf b}^{~}_{ij} \label{GB}
\eea 
where ${\bf b}^\dagger_{ij} = \frac{1}{\sqrt{2}}\left(c^\dagger_{i \uparrow} c^\dagger_{j
  \downarrow} - c^\dagger_{i \downarrow} c^\dagger_{j \uparrow}\right)$
creates a spin singlet on the $i-j$ bond.  $J$ ($>$ 0) is a measure of
singlet or valence bond correlations emphasized by Pauling, i.~e., a
nearest neighbour attraction in the spin singlet channel. In the
present paper we call it as a `bond singlet pairing'(BSP) pseudo
potential. 
The parameter $J$ was chosen as the singlet triplet
splitting in a $2$ site Hubbard model with the same $t$ and $U$, $J =
((U^2 + 16 t^2)^{1/2} - U)/2$. As $U$ becomes larger than the bandwidth this psuedo-potential will become the famous superexchange characteristic of a Mott insulator. 
 As shown in \cite{Baskaran2002}, this
model predicts that undoped graphene is a ``normal'' metal.  The
linearly vanishing density of states at the chemical potential
engenders a critical strength $J_c$ for the BSP to obtain a finite
mean field superconducting $T_c$. The parameter $J$ for graphene
 was less than the critical value, and undoped graphene is not
a superconductor despite Pauling's singlet correlations. Doped
graphene has a finite density of state at the chemical potential and a
superconducting ground state is possible. Black-Schaffer and
Doniach\cite{Schaffer2007} confirmed GB's findings in a detailed and
systematic mean field theory and discovered an important result for
the order parameter symmetry. They found that the lowest energy mean
field solution corresponds to $d + id$ symmetry, an unconventional order
parameter, rather than the extended-$s$ solution. The value of mean field
$T_c$ obtained was an order of magnitude larger than room temperature!

Although results of the mean-field theory are encouraging, it is far
from certain that the superconducting ground state is stable to
quantum fluctuations. In particular, the GB Hamiltonian does not
include $U$ which inhibits local number fluctuations. In the more
basic Hubbard model, a superconducting state will suffer further
quantum mechanical phase fluctuations since $U$ inhibits local number
fluctuations. The key question therefore is does the singlet promotion
arising out of the local correlation physics strong enough to resist
the destruction of superconductivity arising out of quantum phase
fluctuations induced by $U$?

\begin{figure}
\includegraphics[width=\figwidth]{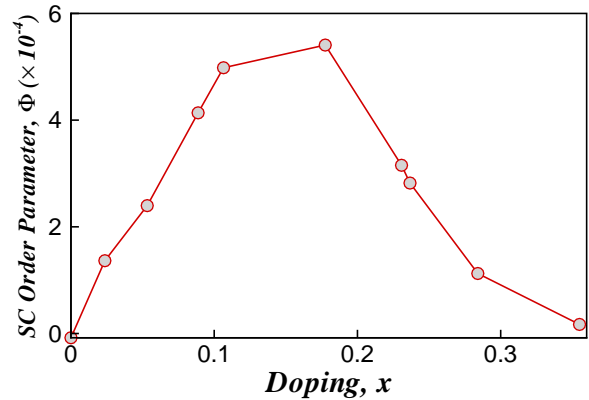}
  \caption{Doping dependence of superconducting order parameter $\Phi$ as
    obtained from VMC calculation of the Hubbard model on a honeycomb
    lattice for $U/t=2.4$.}
  \label{VMCSCOP}
\end{figure}

To investigate the possibility of superconductivity in doped graphene
we construct a variational ground state and optimize it using
variational quantum Monte Carlo (VMC).\cite{Ceperley1977} The ground state we construct is motivated by the mean-field theory of the GB model
\bea
{\cal H}_{GB}^{MF} & = & \sum_{\bk} \left( -t \varepsilon(\bk) (a^\dagger_{\bk \sigma} b_{\bk \sigma})  + \mbox{h.~c.} \right) \non \\ && - \mu_f \sum_{\bk}  \left(n^a_{\bk \sigma} + n^b_{\bk \sigma} \right) \label{HMFk} \\
&& -\sum_{\bk} \left( \Delta(\bk) (a^\dagger_{\bk \uparrow} b^\dagger_{-\bk \downarrow} - a^\dagger_{\bk \downarrow} b^\dagger_{-\bk \uparrow} ) + \mbox{h.~c.} 
\right) \non
\eea
where $a_{\bk \sigma}, b_{\bk \sigma}$ are electron operators on the A
and B sublattices, $\bk$ runs over the Brillouin zone of the
triangular Bravais lattice, $\ba_\alpha, \alpha=1,2,3$ are vectors
that go from an A site to the three nearest B sublattice sites. The
free electronic dispersion is determined by the function
$\varepsilon(\bk) = \sum_{\alpha} e^{i \bk \cdot \ba_\alpha}$, and the
superconducting gap function $\Delta(\bk) = \sum_{\alpha} \Delta_{\alpha} e^{i
  \bk \cdot \ba_\alpha}$. The $d + id$ symmetry motivated by the
meanfield solution\cite{Schaffer2007} provides $\Delta_{\alpha} = \Delta
e^{\frac{i 2 \pi (\alpha -1)}{3}}$ where $\Delta$ is the ``gap parameter'', $\mu_f$ is a ``Hartree shift''. Starting from this mean field theory, we construct a BCS state $\ket{BCS}_N$ with an appropriate number $N$ of electrons. If we work with a lattice with $L$ sites, this corresponds to a hole doping of  $1 - N/L$. Our candidate ground state $\ket{\Psi}$ is now a state with a Gutzwiller-Jastrow factor\cite{Gutzwiller1965,Shiba1989} $g$
\bea
\ket{\Psi} = g^{{\cal D}}\ket{BCS}_N \label{VWF}
\eea
where ${\cal D} = \sum_{i} (n^a_{i \uparrow}n^a_{i \downarrow} +n^b_{i \uparrow}n^b_{i \downarrow}  )$ is the operator that
counts the number of doubly occupied sites. The wavefunction \prn{VWF}
with partial Gutzwiller projection has three variational parameters:
the gap parameter $\Delta$, the Hartree shift $\mu_f$, and the Gutzwiller-Jastrow
factor $g$. The ground state energy $\bra{\Psi} {\cal H}_H \ket{\Psi}$ is
calculated using quantum Monte Carlo method\cite{Ceperley1977}, and is
optimized with respect to the variational parameters. Details
regarding construction and optimization of the wavefunction can be
found in the methods section.

\begin{figure}
\includegraphics[width=\figwidth]{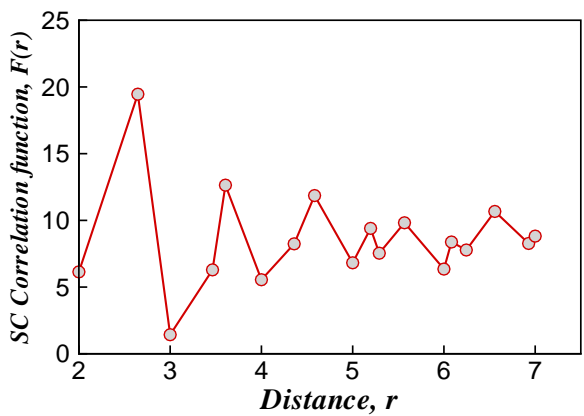}
\caption{Dependence of superconducting correlation function $F(r)$ on
  distance $r$ as obtained from VMC calculation of the Hubbard model
  on a honeycomb lattice for $U/t=2.4, x = 0.2$.}
  \label{SCOPdepr}
\end{figure}

We monitor superconductivity by calculating the following correlation function using the optimized wavefunctions
\bea
F_{\alpha \beta}(\bR_i - \bR_j) = \mean{{\bf b}^\dagger_{i \ba_\alpha} {\bf b}_{j \ba_\beta}}
\eea
where ${\bf b}^\dagger_{i \ba_\alpha}$ is the {\it electron} singlet operator that creates a singlet between the $A$ site in the $i$-th unit cell and the $B$ site connected to it by the vector $\ba_\alpha$ (this is just ${\bf b}^\dagger_{ij}$ defined earlier with a minor change of notation).  The superconducting order parameter, off-diagonal long range order (ODLRO), is
\bea
\Phi = \lim_{|\bR_i - \bR_j| \longrightarrow \infty}F(\bR_i - \bR_j) .
\eea
where $F(\bR_i - \bR_j) = \sum_{\alpha} F_{\alpha \alpha}(\bR_i - \bR_j)$. All results we show in this paper are performed on lattices with $13^2$ unit cells.

The superconducting order parameter $\Phi$ as a function of doping,
calculated for physical parameters corresponding to graphene, obtained
using the optimized wavefunction is shown in
\fig{VMCSCOP}. Remarkably, a ``superconducting dome'', reminiscent of
cuprates,\cite{Paramekanti2004} is obtained and is consistent with the RVB physics.
The result indicates that undoped graphene had no long range
superconducting order consistent with physical arguments and
mean-field theory\cite{Schaffer2007} of the phenomenological GB
Hamiltonian. Interestingly, the present calculation suggests an
``optimal doping'' $x$ of about 0.2 at which the the ODLRO attains a maximum. These calculations strongly suggest a superconducting ground state in doped graphene.

We now further investigate the system near optimal doping in order to
estimate $T_c$.  \Fig{SCOPdepr} shows a plot of the order parameter
function $F(r)$ as function of the separation $r$. The function has
oscillations up to about six to seven lattice spacings and then attains a nearly
constant value. From an exponential fit one can infer that the coherence length
$\xi$ of the superconductor is about six to seven lattice
spacings. A crude estimate of an upper bound of transition temperature can then be obtained by using results from weak coupling BCS theory,
using $k_b T_c = \frac{1}{1.764} \frac{\hbar v_F}{\pi
  \xi}$. \textit{Conservative estimates give us $k_b T_c = \frac{t}{50}$,
i.e., $T_c$ is about twice room temperature.} Evidently, this is an
upper bound,  and an order of magnitude lower than the mean-field
theory estimates of Black-Schaffer and Doniach\cite{Schaffer2007}. Further improvement of  our estimate of $T_c$ becomes technically difficult. It is interesting to compare these results with those
obtained in a Hubbard model on a square lattice that captures cuprate
physics. In this latter case, a similar estimate of the coherence length
$\xi$ is about two to three lattice spacings\cite{Paramekanti2004};
however, the hopping scale is nearly a magnitude lower and the
estimate of $T_c$ is about $2 T_{\rm Room}$. Again, this provides further
support for the possibility of high temperature superconductivity in
graphene.

\begin{figure}
\includegraphics[width=\figwidth]{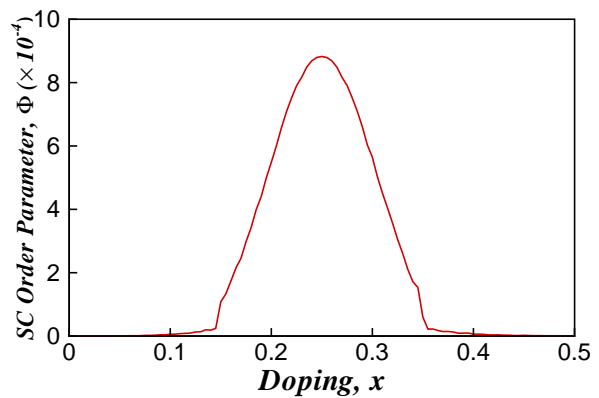}
  \caption{Doping dependence of superconducting order parameter as obtained from Slave rotor mean field theory of the $t$-$J$-$U$ model.} 
  \label{SRSCOP}
\end{figure}

It is interesting to see if the results of the VMC can be reproduced
in a ``simple mean-field type'' analysis that accounts for the
presence of $U$. The recently developed slave rotor\cite{Florens2004}
technique is useful to study an intermediate coupling regime that
allows for number fluctuations at a site, and has been used
successfully, in the context of cuprates, to study the $t$-$J$-$U$
model.\cite{Zhao2007} We adopt a similar approach and use a
$t$-$J$-$U$ model which is equivalent to introducing a Hubbard repulsion $U$
in the GB Hamiltonian. The details of the slave rotor technique is
presented in the methods section. The ODLRO calculated from the slave
rotor analysis is shown in \Fig{SRSCOP}, and bears a remarkable
qualitative resemblance with the VMC result, i.~e., there is an
optimal doping that produces superconductivity.  This result, again, supports the possibility of high temperature
superconductivity in doped graphene. 

It is important to ask about the possibility of competing orders that
could overshadow superconductivity at optimal doping that we have
found. \citebyname{Honerkamp2008} has addressed this issue by means of a
functional renormalization group study of a general Hamiltonian on the
honeycomb lattice. He finds that in the regime of physical parameter
corresponding to graphene, the system appears to flow towards a $d+id$
superconducting state as the temperature is lowered.

It is interesting to contrast the superconductivity in cuprates with
that in graphene.  In the case of cuprates (doped Mott insulators),
Bloch electrons in the entire Brillouin zone are affected by the
Hubbard $U$ which is larger than the bandwidth. To this extent all
electrons participate in superexchange or singlet bond
formation. Consequently correlation hole development is complete,
i.~e., an electron at a given site with an up spin manages to avoid an
electron with down spin on its site completely, at low energy scales.
Whereas, in the broad band graphene, only those Bloch electrons in the
range of energy scale of $U$ around chemical potential are affected by the
on site repulsion. Since this scale is about half the bandwidth, about
one half of the electrons are involved in singlet bond formation in
the ground and low energy states. Consequently correlation hole
development is not complete. The key point is that there is a
sufficiently enhanced singlet correlation, compared to free Fermi gas,
to be able to support superconductivity induced by the on site Coulomb
repulsions in optimally doped graphene. This heuristic picture is
supported by the variational Monte Carlo result and the slave rotor
analysis that we have presented in this paper.

Our prediction of high temperature superconductivity raises some
obvious questions. Intercalated graphite can be viewed as a set of doped graphene layers
that have a strong 3 dimensional electronic coupling. Maximum $T_c$
obtained in these systems is around 16 K.\cite{Klapwijk2005,Weller2005} Systems
such as CaC$_{6}$ has a doping close to optimal doping that we have
calculated. Why is the $T_c$ so low?  On the other hand,
superconducting signals with a $T_c$ around 60 K and higher have been
reported in the past in pyrolitic graphite containing
sulfur.\cite{Silva2001,Kopelevich2007} A closer inspection reveals that for systems
like CaC$_{6}$ i) an enhanced 3 dimensionality arising through the
intercalant orbitals makes the effect of Hubbard $U$ less important
(effect of $U$ for a given bandwidth progressively becomes important
as we go down in dimensions) and ii) encouragement of charge density
wave order arising from the intercalant order. Sulfur doped graphite,
however, gives a hope that there is a possibility of high temperature
superconductivity. Our present theoretical prediction should encourage
experimentalists to study graphite from superconductivity point of
view systematically, along the line pioneered by Kopelevich and
collaborators.\cite{Kopelevich2007} In the past there has been claims
(unfortunately not reproducible) of Josephson like signals in graphite
and carbon based materials;\cite{Lebedev2008}  Again, our result should encourage
revival of studies along these lines.

Simple doping of a freely hanging graphene layer by gate control to
the desired optimal doping of 10 -- 20 \% is not 
experimentally feasible at the present moment. It will be interesting to
discover experimental methods that will allow us to attain these higher
doping values. A simple estimate shows that a large cohesion energy arising
from the strong $\sigma$ bond that stabilizes the honeycomb structure
will maintain the structural integrity of graphene.

The discovery of time reversal symmetry braking $d + id$ order\cite{Schaffer2007} for the
superconducting state, within our RVB mechanism is very
interesting. This unconventional order parameter has its own
signatures in several physical properties: i) spontaneous currents in
domain walls, ii) chiral domain wall states iii) unusual vortex
structure and iv) large magnetic fields arising from the $d=2$ angular
momentum of the cooper pairs, which could be detected $\mu$SR
measurements. Suggestions for experimental determination of such an
order by means of Andreev conductance spectra have been made by
\citebyname{Jiang2008}

There are also several theoretical and experimental issues that needs to be addressed. 
It is known that graphene realized in experimental systems contains, adsorbed species, inhomogeneities, curvature, ripples etc.\cite{Fasolino2007} Is the superconducting ground state stable to these ``perturbations''? In particular our theory
gives a substantial ODLRO even for small doping. If disorder effects are indeed suppressing a fragile Kosterlitz-Thouless order in the currently available doping regime in real systems, 
one could uncover the hidden superconductivity by disorder control or study of cooper pair fluctuation effects. Further analysis is necessary to address these issues.

\section{Methods}
\subsection{VMC}
Diagonalizing the kinetic energy part of ~\eqn{HMFk} results in two
bands $c^{\dagger}_{\bk\sigma}$ and $d^{\dagger}_{\bk\sigma}$. The
superconducting pairing term splits into intra-band pairing and
inter-band pairing. The latter being unimportant at zero temperature
can be dropped giving 
\bea
{\cal H}_{MF} & = & \sum_{\bk} \left(E^+(\bk) c^\dagger_{\bk \sigma} c_{\bk \sigma} + E^-(\bk) d^\dagger_{\bk \sigma} d_{\bk \sigma}\right) \non \\
             & + & \sum_{\bk} \left(\Delta_d(\bk) \left( c^\dagger_{\bk \uparrow} c^\dagger_{-\bk \downarrow}  -  d^\dagger_{\bk \uparrow} d^\dagger_{-\bk \downarrow}  \right) + \mbox{h.~c.}\right) \non
\eea
where, $\Delta_d(\bk) = \sum_{\alpha}\Delta_{\alpha}
\cos{(\bk\cdot\ba_{\alpha}-\varphi(\bk))}$, $\varphi(\bk) = \arg{\varepsilon(\bk)}$. Thus, the
variational ground state  can be written as $\ket{BCS} =
\prod_{\bk,\alpha=\pm} (u^{\alpha}_{\bk} +
v^{\alpha}_{\bk}e^\dagger_{\bk \alpha \uparrow} e^\dagger_{-\bk \alpha
  \downarrow})\ket{0}$ where $e^\dagger_{\bk + \sigma} = c^\dagger_{\bk \sigma}$,  $e^\dagger_{\bk - \sigma} = d^\dagger_{\bk \sigma}$ and the BCS
coherence factors $\frac{v^{\alpha}_{\bk}}{u^{\alpha}_{\bk}}= -\alpha \frac{\Delta_d (\bk)}{E^{\alpha}_{\bk} +
  \sqrt{ (E^{\alpha}_{\bk})^2 + |\Delta_d(\bk)|^2 }}$.

To deal with system containing fixed number of particles, the
variational wavefunction is projected onto a fixed number subspace -
$\ket{BCS}_N$. The trial wavefunction is obtained by applying the
Jastrow factor to this state - $g^{{\cal D}} \ket{BCS}_N$. Energy of this state is calculated using Monte-Carlo sampling,  which is then optimized using the
simplex method.

\subsection{SlaveRotor}
The $t$-$J$-$U$ model can be written as ${\cal H} = -
\sum_{ij\sigma}t^{}_{ij} c^\dagger_{i\sigma} c^{}_{j\sigma} + U \sum_i
n^{}_{i\uparrow}n^{}_{i\downarrow} +
J\sum_{\mean{ij}}\bS_i\cdot\bS_j$. The idea is to decompose electronic
degree of freedom into spinon and charge (rotor) degrees of freedom
(DOFs) - $c_{i\sigma}^{\dagger} = f_{i\sigma}^{\dagger}e^{-i\theta_i}$
where $f_{i\sigma}^{\dagger}$ creates a spin $\sigma$ at site $i$ and
$e^{\pm i\theta_i}$ are the ladder operators for electron density,
$n$.

In this representation, $t$-$J$-$U$ model takes the form - ${\cal
  H}_{SR} = - \sum_{ij\sigma}t^{}_{ij} f^\dagger_{i\sigma}
f^{}_{j\sigma}e^{-i\left(\theta_i-\theta_j\right)} + \frac{U}{2}
\sum_i n^{\theta}_i(n^{\theta}_i-1) +
J\sum_{\mean{ij}}\bS_i^f\cdot\bS_j^f$. The total number of particles
at any site has to be unity to remove unphysical states in the
expanded Hilbert space.  At mean field level, the spinon (rotor) DOF
can be integrated out to give a Hamiltonian in the rotor (spinon)
space. The two Hamiltonians are coupled via the kinetic energy
term. They are solved self-consistently using standard
techniques\cite{Zhao2007}.

\subsection*{Acknowledgement}
SP and VBS thank SERC, DST for support of this work. VBS acknowledges generous support from DST via a Ramanujan grant.


\begin{thebibliography}{30}
\expandafter\ifx\csname natexlab\endcsname\relax\def\natexlab#1{#1}\fi
\expandafter\ifx\csname url\endcsname\relax
  \def\url#1{\texttt{#1}}\fi
\expandafter\ifx\csname urlprefix\endcsname\relax\def\urlprefix{URL }\fi

\bibitem[{Bednorz \& Mueller(1986)}]{Bednorz1986}
Bednorz, J. \& Mueller, K.
\newblock Possible high T$_c$ superconductivity in the Ba-La-Cu-O system.
\newblock \emph{Z. Phys. B} \textbf{64}, 189 (1986).

\bibitem[{Anderson(1987)}]{Anderson1987}
Anderson, P.
\newblock The Resonating Valence Bond State in La$_2$CuO$_4$ and
  Superconductivity.
\newblock \emph{Science} \textbf{235}, 1196 (1987).

\bibitem[{Nagamatsu \emph{et~al.}(2001)Nagamatsu, Nakagawa, Muranaka, Zenitani
  \& Akimitsu}]{Nagamatsu2001}
Nagamatsu, J., Nakagawa, N., Muranaka, T., Zenitani, Y. \& Akimitsu, J.
\newblock Superconductivity at 39 K in magnesium diboride.
\newblock \emph{Nature} \textbf{410}, 63 (2001).

\bibitem[{Kamihara \emph{et~al.}(2008)Kamihara, Watanabe, Hirano \&
  Hosono}]{Kamihara2008}
Kamihara, Y., Watanabe, T., Hirano, M. \& Hosono, H.
\newblock \protect{Iron-Based Layered Superconductor LaO$_{1-x}$F$_x$FeAs $(x =
  0.05-0.12)$ with $T_c$ = 26}.
\newblock \emph{J. Am. Chem. Soc.} \textbf{130}, 3296--3297 (2008).

\bibitem[{Geim \& Novoselov(2007)}]{Geim2007}
Geim, A.~K. \& Novoselov, K.~S.
\newblock The Rise of Graphene.
\newblock \emph{Nat. Mater.} \textbf{6}, 183--191 (2007).

\bibitem[{Katsnelson(2007)}]{Katsnelson2007}
Katsnelson, M.~I.
\newblock Graphene: carbon in two dimensions.
\newblock \emph{Materials Today} \textbf{10}, 20--27 (2007).

\bibitem[{Castro~Neto \emph{et~al.}(2008)Castro~Neto, Guinea, Peres, Novoselov
  \& Geim}]{CastroNeto2007}
Castro~Neto, A.~H., Guinea, F., Peres, N.~M.~R., Novoselov, K.~S. \& Geim,
  A.~K.
\newblock The electronic properties of graphene.
\newblock \emph{arXiv.org:0709.1163v2}  (2008).

\bibitem[{Baskaran(2002)}]{Baskaran2002}
Baskaran, G.
\newblock Resonating-valence-bond contribution to superconductivity in MgB$_2$.
\newblock \emph{Phys. Rev. B} \textbf{65}, 212505 (2002).

\bibitem[{Black-Schaffer \& Doniach(2007)}]{Schaffer2007}
Black-Schaffer, A.~M. \& Doniach, S.
\newblock Resonating valence bonds and mean-field d-wave superconductivity in
  graphite.
\newblock \emph{Phys. Rev. B} \textbf{75}, 134512 (2007).

\bibitem[{Uchoa \& Neto(2007)}]{Uchoa2007}
Uchoa, B. \& Neto, A. H.~C.
\newblock Superconducting States of Pure and Doped Graphene.
\newblock \emph{Phys. Rev. Lett.} \textbf{98}, 146801 (2007).

\bibitem[{Choy \& McKinnon(1995)}]{Choy1995}
Choy, T.~C. \& McKinnon, B.~A.
\newblock Significance of nonorthogonality in tight-binding models. II. The
  possibility of high-$Tc$ superconductivity in intercalation compounds.
\newblock \emph{Phys. Rev. B} \textbf{52}, 14539--14543 (1995).

\bibitem[{Furukawa(2001)}]{Furukawa2001}
Furukawa, N.
\newblock Antiferromagnetism of the Hubbard Model on a Layered Honeycomb
  Lattice — Is MgB$_2$ a Nearly-Antiferromagnetic Metal?
\newblock \emph{J. Phys. Soc. Jpn.} \textbf{70}, 1483 (2001).

\bibitem[{Onari \emph{et~al.}(2003)Onari, Arita, Kuroki \& Aoki}]{Onari2003}
Onari, S., Arita, R., Kuroki, K. \& Aoki, H.
\newblock Superconductivity in repulsive electron systems with
  three-dimensional disconnected Fermi surfaces.
\newblock \emph{Phys. Rev. B} \textbf{68}, 024525 (2003).

\bibitem[{Baeriswyl \& Jackelman(1995)}]{Hubbard1995}
Baeriswyl, D. \& Jackelman, E.
\newblock In \emph{The Hubbard Model: Its Physics and Mathematical Physics}
  (ed. Baeriswyl, D.), 393 (Plenum, New York, 1995).

\bibitem[{Baskaran \& Jafari(2002)}]{BaskaranJafari2002}
Baskaran, G. \& Jafari, S.~A.
\newblock Gapless Spin-1 Neutral Collective Mode Branch for Graphite.
\newblock \emph{Phys. Rev. Lett.} \textbf{89}, 016402 (2002).

\bibitem[{Pauling(1960)}]{Pauling1960}
Pauling, L.
\newblock \emph{Nature of the chemical bond} (Cornell University Presss, NY,
  1960).

\bibitem[{Ceperley \emph{et~al.}(1977)Ceperley, Chester \&
  Kalos}]{Ceperley1977}
Ceperley, D., Chester, G. \& Kalos, M.
\newblock Monte Carlo simulation of a many-fermion study.
\newblock \emph{Phys. Rev. B} \textbf{16}, 3081 (1977).

\bibitem[{Gutzwiller(1965)}]{Gutzwiller1965}
Gutzwiller, M.~C.
\newblock Correlated electrons in a narrow $s$ band.
\newblock \emph{Phys. Rev.} \textbf{137}, A1726--1735 (1965).

\bibitem[{Shiba(1989)}]{Shiba1989}
Shiba, H.
\newblock \emph{Two-Dimensional Strongly Correlated Electron Systems, Zi-zhao
  Gan and Zhao-bin Su, Ed.}, chap. Some aspects of strongly correlated
  electronic systems -- variational Monte Carlo studies (Gordon and Breach
  Science Publishers, 1989).

\bibitem[{Paramekanti \emph{et~al.}(2004)Paramekanti, Randeria \&
  Trivedi}]{Paramekanti2004}
Paramekanti, A., Randeria, M. \& Trivedi, N.
\newblock {High-Tc superconductors: A variational theory of the superconducting
  state}.
\newblock \emph{Phys. Rev. B} \textbf{70}, 054504 (2004).

\bibitem[{Florens \& Georges(2004)}]{Florens2004}
Florens, S. \& Georges, A.
\newblock Slave-rotor mean-field theories of strongly correlated systems and
  the Mott transitionin finite dimensions.
\newblock \emph{Phys. Rev. B} \textbf{70}, 035114 (2004).

\bibitem[{Zhao \& Paramekanti(2007)}]{Zhao2007}
Zhao, E. \& Paramekanti, A.
\newblock Self-consistent slave rotor mean-field theory for strongly correlated
  systems.
\newblock \emph{Phys. Rev. B} \textbf{76}, 195101 (2007).

\bibitem[{Honerkamp(2008)}]{Honerkamp2008}
Honerkamp, C.
\newblock Density Waves and Cooper Pairing on the Honeycomb Lattice.
\newblock \emph{Phys. Rev. Lett.} \textbf{100}, 146404 (2008).

\bibitem[{Klapwijk(2005)}]{Klapwijk2005}
Klapwijk, T.~M.
\newblock Superconductivity: Would your graphite pencil superconduct?
\newblock \emph{Nat. Phys.} \textbf{1}, 17 (2005).

\bibitem[{Weller \emph{et~al.}(2005)Weller, Ellerby, Saxena, Smith \&
  Skipper}]{Weller2005}
Weller, T.~E., Ellerby, M., Saxena, S.~S., Smith, R.~P. \& Skipper, N.~T.
\newblock Superconductivity in the intercalated graphite compounds C$_6$Yb and
  C$_6$Ca.
\newblock \emph{Nat. Phys.} \textbf{1}, 39 (2005).

\bibitem[{da~Silva \emph{et~al.}(2001)da~Silva, Torres \&
  Kopelevich}]{Silva2001}
da~Silva, R.~R., Torres, J. H.~S. \& Kopelevich, Y.
\newblock Indication of Superconductivity at 35 K in Graphite-Sulfur
  Composites.
\newblock \emph{Phys. Rev. Lett.} \textbf{87}, 147001 (2001).

\bibitem[{Kopelevich \& Esquinazi(2007)}]{Kopelevich2007}
Kopelevich, Y. \& Esquinazi, P.
\newblock Ferromagnetism and Superconductivity in Carbon-based Systems.
\newblock \emph{J. Low Temp. Phys.} \textbf{146}, 629 (2007).

\bibitem[{Lebedev(2008)}]{Lebedev2008}
Lebedev, S.
\newblock Search for the Reasons of Josephson Like Behavior of Thin Granular
  Carbon Films (2008).
\newblock Cond-mat/0802.4197, and references therein.

\bibitem[{Jiang \emph{et~al.}(2008)Jiang, Yao, Carlson, Chen \& Hu}]{Jiang2008}
Jiang, Y., Yao, D.-X., Carlson, E.~W., Chen, H.-D. \& Hu, J.
\newblock Andreev conductance in the $d+id$-wave superconducting states of
  graphene.
\newblock \emph{Phys. Rev. B} \textbf{77}, 235420 (2008).

\bibitem[{Fasolino \emph{et~al.}(2007)Fasolino, Los \&
  Katsnelson}]{Fasolino2007}
Fasolino, A., Los, J. \& Katsnelson, M.
\newblock Intrinsic ripples in graphene.
\newblock \emph{Nat. Mat.} \textbf{6}, 858 (2007).

\end{thebibliography}

\end{document}